# Managing mental & psychological wellbeing amidst COVID-19 pandemic: Positive psychology interventions

Maria Tresita Paul V.[1] & Uma Devi N.[2]

[1 & 2] *Bharathiar School of Management and Entrepreneur Development (BSMED), Bharathiar University, Tamil Nadu, India.*
*maria.tresi@gmail.com*

**ABSTRACT**

COVID-19 pandemic has shaken the roots of healthcare facilities worldwide, with the US being one of the most affected countries irrespective of being a superpower. Along with the current pandemic, COVID-19 can cause a secondary crisis of mental health pandemic if left unignored. Various studies from past epidemics, financial turmoil and pandemic, especially SARS and MERS, have shown a steep increase in mental and psychological issues like depression, low quality of life, self-harm and suicidal tendencies among general populations. The most venerable being the individuals infected and cured due to social discrimination. The government is taking steps to contain and prevent further infections of COVID 19. However, the mental and psychological wellbeing of people is still left ignored in developing countries like India. There is a significant gap in India concerning mental and psychological health still being stigmatized and considered 'non-existent'. This study's effort is to highlight the importance of mental and psychological health and to suggest interventions based on positive psychology literature. These interventions can support the wellbeing of people acting as a psychological first aid.

## I. INTRODUCTION

China reported the first documented case of newly identified chronic respiratory illness COVID-19, on December 16th, 2019, in Wuhan province. Unaware of the upcoming global catastrophe, by this time, the rest of the world was in a celebration mode preparing for the new year 2020. Diseased doctor 'Li Wenliang' was the whistleblower, who alerted about this suspicious new disease novel coronavirus - COVID-19 via social media. The Chinese government's numerous attempts to suppress this caught the attention of international media. By late December and early January of 2020, reports of confirmed cases of COVID 19 diseases spreading outside of China to countries like Americal, Italy, England & India, came to light affirming the human to human transition. With the global news covering COVID-19, came enormous pieces of information causing anxiety, distress and fear worldwide among people, of this unknown new disease infecting and killing thousands worldwide (Wang et al. 2020), especially the vulnerable and elderly (Centers for Disease Control and Prevention 2020). By January 2020, the World Health Organisation (WHO) announced COVID-19 epidemic outbreak, a public health emergency of international significance, and reported a high risk of COVID-19 spreading to other countries around the world. The COVID-19 virus is a zoonotic infection thought to have originated from pangolins, snakes and bats in wet markets of Wuhan (Ji et al. 2020). By January 2020 WHO confirmed the human-to-human transition of COVID 19. In March 2020, the WHO assessed and declared coronavirus as a pandemic. Between March to June 2020, there is an exponential growth of coronavirus disease infected victims. Pandemic related containment measures worldwide as recommended by WHO are 'quarantine, social distancing, and self-isolation'.

Recent research has shown that a long period of 'quarantine, social distancing, and self-isolation' in already uncertain situations like pandemic can harm mental and psychological wellbeing globally (Brooks et al., 2020; Dubey et al., 2020; Qui et al., 2020). The need of mental wellbeing and stress coping of medical practitioners has become crucial area of study (Paul et al., 2021). Positive psychology has proven to heal and enhance the mental and psychological wellbeing of individuals (Seligman 2004; Seligman and Csikszentmihalyi 2014; Slade 2010; Vázquez et al., 2009).
Positive psychology is explained as the scientific study of "what makes life most worth living" (Peterson, 2008), focusing on a) positive experiences, (e.g. joy, happiness, life satisfaction, inspiration and love); b) positive traits and states (e.g. resilience, optimism, gratitude, hope, efficacy and compassion); c) positive institutions (applying positive principles within institutions). Positive psychology interventions have been used for decades to enhance the mental and psychological wellbeing of people (Bolier et al., 2013; Gander, 2016; Sin and Lyubomirsky, 2009; Pawelski, 2020). The present study examines and contributes to the ongoing research on the mental and psychological impact of COVID-19 in two ways. First, the study examines the possible mental and psychological consequences of COVID-19 among the general population, adding to the existing literature on COVID-19.

**Fig 1: New cases of COVID-19 overtime worldwide**

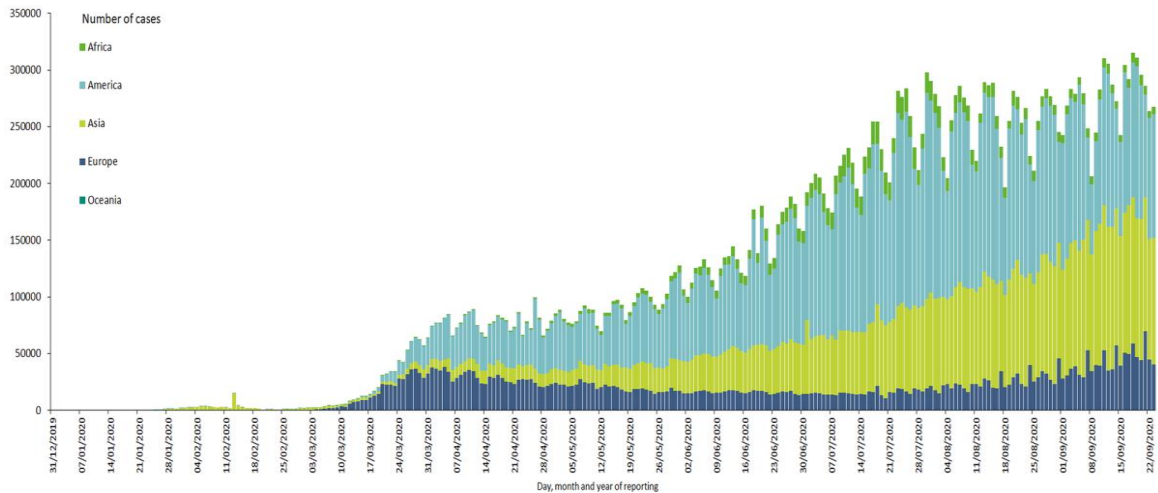

affected a university medical student, had a travel history from Wuhan, China. Although the pandemic affected India by late January, the fear due to constant unverified information via various social media and worldwide web had already affected and engulfed Indians by late December and early January. Pandemics have an intense mental, and psychological toll on people, mainly by triggering "the constant fear of getting infected by an unpredictable virus that doctors don't understand and are too much for them to handle". The pandemics are not just medical phenomenon. Pandemic disrupts the professional and personal lives of people and the economic and social environment across countries at multiple levels. As on September 24$^{th}$ in India, there are 5,646,010 confirmed cases of COVID-19 including 90020 fatalities, as per the latest data provided by the World Health Organization (WHO). The critical strategies adopted to contain the outbreaks of this nature are physical distancing or social distancing and isolation. Both can have significant impacts on the lives and relationships of people facing the pandemic. The Indian has undertaken these significant preventive measures to tackle COVID-19.

**Table 1:** Janta Curfew/Lockdown-Unlock phases & COVID cases

| | Phase | From | To | Days | Number of infections*# | Daily case increase* | Total number of death* |
|---|---|---|---|---|---|---|---|
| Lockdown | I | 25.03.2020 | 14.04.2020 | 21 | 10363 | 1211 | 383 |
| | II | 15.04.2020 | 03.05.2020 | 19 | 39980 | 2644 | 1301 |
| | III | 04.05.2020 | 17.05.2020 | 14 | 90927 | 4987 | 2872 |
| | IV | 18.05.2020 | 31.05.2020 | 14 | 182143 | 8380 | 5164 |
| Unlock | I | 01.06.2020 | 30.06.2020 | 30 | 566840 | 18522 | 16893 |
| | II | 01.07.2020 | 31.07.2020 | 31 | 1638870 | 55078 | 35747 |
| | III | 01.08.2020 | 31.08.2020 | 31 | 3621245 | 78512 | 64469 |
| | IV | 01.09.2020 | 30.09.2020$ | 30 | 5818570 | 86508 | 92290 |

**Source:** Author (data from WHO). *As on the end day. #Cumulative total. $ As on 24.09.2020.

**Social Distancing & Isolation: Janta Curfew :** In India, the government initiated the 'Janta Curfew', which is self-imposed complete lockdown of the country. On March 22 2020, Narendra Modi, Prime Minister of India, urged citizens to stay in their home, avoiding going out. The Jantha curfew was imposed to save oneself from getting infected also to protect others from being exposed to COVID-19. Lockdown was implmented in four phases, followed by gradual four stages of unlocking mechanisim, yet in process. Table 1 depicts that there is a steep increase in the number of COVID-19 infected case, death rates and the daily increase in cases. Those working in emergency services like medical staff, police, sanitary workers, firefighters, media and army personnel who are serving the nation are exempted from participating in Jantha curfew. India has seen four phases of Jantha curfew, and the fifth phase with slight relaxations (for those states where corona cases have come to control) is running at present. Like any other fast-spreading infection, COVID-19 comes with an exponentially increasing barrage of misinformation thrown continuously at us via the worldwide web, social media, fuelling stress and mass hysteria. Besides, the 'fear of transmission' begets stigma, marginalization and xenophobia, kicking in the 'fear of fellow humans'.

**AarogyaSetu – Application :** AarogyaSetu is a mobile application developed by the Government of India to connect essential health services with the citizens of India in the fight against coronavirus pandemic. The application is aimed at supplementing the initiatives of the Government of India, especially the Department of Health, in proactively informing and reaching out the users of the application concerning hazards, healthy practices and relevant advisories about the containment of COVID-19. This app is tracking software, which uses the smartphone's Bluetooth and GPS features to track COVID-19 infections. This app is made available to users of both android and IOS operating system mobiles. It is freely available to be downloaded from google play store and government website (https://www.mygov.in/covid-19). The app is developed to determine if there is a risk of getting in contact (within six feet) to a coronavirus infected patient. Using the location data retrieved from GPS, it also predicts if a person is in from any of the infected zones declared by the government. The app is an upgraded version of the Corona Kavach application, which was discontinued by the Indian government.

## II. LITERATURE REVIEW

**Mental health impact of COVID-19 :** In the light of COVID-19's recent and abrupt emergence, research in this field is understandably constrained. However, there is a strong opinion among researchers, psychologist and social scientist across the world for the need of quality research, understanding the underlying mechanism and effects of coronavirus disease on the mental health of populations. Further, there is an immediate need for researchers to study the impact of constant consumption of information and rumours regarding COVID-19 and the mental health consequences of it among people and how these adverse effects can be mitigated in the pandemic environment (Holmes et al., 2020). During the initial COVID-19 outbreak several studies from China reported associations of COVID-19 with increased anxiety, depression and stress (Cao et al., 2020; Wang et al., 2020; Zhang et al., 2020). The study conducted in Iran emphasized the roles of misinformation and isolation, uncertainty, unpredictability and the seriousness of disease as fuel to stress and mental morbidity (Zandifar & Badrfam, 2020). The authors pressed for both strengthening of social capital and better mental health service to people for reducing the psychological impact of COVID-19. Research conducted in Japan revealed the effect of economic consequences of coronavirus on the mental wellbeing of people, high levels of panic and fear-induced behaviours, such as stockpiling and hoarding of resources (Shigemura et al., 2020). Another research in China conducted on people who were isolated revealed a sharp spike in their levels of stress, anxiety and reduced sleep quality (Xiao et al., 2020b). A study conducted in India revealed one-third of their respondents had significant psychological impact due to COVID-19, and it was higher among younger aged people, females and those already suffering from comorbid physical illness (Varshney et al., 2020). The authors highlighted the urgent need of research on systematic & longitudinal psychological need assessment of the general population to help the government in formulation of holistic intervention for affected people. Recent findings emphasize that the negative economic and social consequences of current lockdown, economic downturn and constant exposure to distressing media coverage during COVID-19 contributes to adverse psychological outcomes like reduced social support, increased loneliness, anxiety depression and financial worry among populations (Asmundson & Taylor, 2020; Courtet et al., 2020; Reger et al., 2020).

**Positive Psychology :** Positive psychology is the scientific study of harnessing the power of positive aspects of human life, (e.g., wellbeing and flourishing) to shift individual's perspective in many of their everyday behaviours towards maximizing the potential of their happiness. Positive psychology is a set of personal resources, and principles practising which a person can attain life satisfaction. Martin Seligman defined positive psychology as 'scientific study for optimal human functioning, aiming to discover and promote the aspects that allow individuals as well as communities to thrive' (2005). Positive psychology enables an individual to experience a meaningful & purposeful life, moving from surviving to flourishing. Theorist, psychologists and researchers in the field have proposed and tested positive psychology interventions for improving life-satisfaction and wellbeing. Positive psychology advocates experiencing real happiness rather than superficial happiness. Studies on positive psychology reveal that spending money on experiences boosts happiness even more than spending money on physical belongings (Howell & Hill, 2009). Practising gratitude is a significant contributor to life's satisfaction and shows that the more appreciation individuals cultivate, the happier they are (Seligman et al., 2005). Oxytocin induces greater confidence, empathy and morality in humans, which ensures that showing affection to loved ones gives a significant boost the overall wellbeing (Barraza & Zak, 2009). Individuals who intentionally nurture a positive mood to match the outward emotion they need to show (i.e., in emotional labour) benefit from a more real experience of positive feeling. Conversely, "putting on a happy face" doesn't necessarily make an individual feel happier, but putting in a bit of effort will probably make individuals feel happier (Scott & Barnes, 2011). Individual's happiness is transmissible, and those with happy friends and significant others are more probable to experience happiness in future (Fowler & Christakis, 2008). Individuals who perform acts of kindness to others not only receive a boost in their wellbeing but are also are appreciated by their peers (Layous, et al., 2012). Volunteering time for a cause an individual believes improves their wellbeing and life satisfaction, and can even reduce their depression symptoms (Jenkinson et al., 2013). Spending money on others contributes (e.g. charity) to greater happiness for the giver. (Dunn et al., 2008).

### III. MENTAL AND PSYCHOLOGICAL RED FLAGS: 1ST WAVE COVID-19

The healthcare sector across various countries, including developed nations like the USA, US and Italy, are already overwhelmed. With the exponential rise of COVID-19 infected cases worldwide, doctors and nurses are burnt out working overtime. Healthcare professionals are at constant high risk of getting exposed and infected to COVID-19. People across the globe are competing for even basic medical amenities; this situation gets worse in developing and undeveloped countries. There is a boom of self-proclaimed doctors, claiming to have found the cure, wrong treatments, and unorthodox medical preventive practices. People tend to feel nervous and uncomfortable when there is a sudden drastic change in the environment. In the case of outbreaks of infectious diseases, where the origin or development of the disease and the result are uncertain, misinformation and rumours grow causing a surge in closeminded attitudes among people (Ren et al. 2020).

**Fig 2: Cases of COVID-19 among ten most infected countries**

**Source:** World health Organisation (as on 25.09.2020)

People experience mental health fallouts like confusion, health anxiety, stress, insomnia, panic attacks, loneliness, depression, suicidal tendencies. A pandemic affects not just physical health but more importantly, mental health and wellbeing of millions of people (Brooks et al., 2020; Shigemura et al., 2019). The current pandemic is challenging the priorities and agenda of healthcare professionals, bringing psychiatrists and other mental health professionals to spotlight (Yu-Tao et al., 2019). There are direct and indirect adverse longterm effects of COVID 19 pandemic on the mental and psychological wellbeing of people in the following ways:

- *Anxiety and fear* of losing employment, business and livelihood due to economic shutdown, isolation, closed markets, social setup changes and travel limitations (For MSMEs, tourism)
- A growing feeling of *insecurity* to fulfil the needs of the family.

- *Fear* of going outside the home owing to the extreme rate of infections.
- *Stress and depression* caused due to social distancing from family members.
- *Compulsive tendency to hoard* food, medical resources and groceries out of fear of non-availability.
- Rise of *domestic violence* cases against women.
- Increase in *child sexual and physical abuse* incidents.
- *Social discrimination* of infected and cured patients and their families.
- *Fear of being discriminated* if show symptoms of fever, cold, cough, even though simple flu.
- The *financial stress* of the hefty medical bills, in case, get infected.
- *Psychological distress* causing a *spiral of stress* which grows and is passed on to others much time owing to *constant negative reinforcing* news in media (Growth in the number of Coronavirus infection and death cases).
- *Extreme burnout, sense of isolation, frustration, fatigue, fear of infection or constant guilt* of transmitting COVID 19 infection to others among medical care workers, humanitarians, police and media personnel.
- *The trauma and distress due to constant fear of reliving a health crisis* for people who have once experienced infectious disease epidemic or pandemic in their lifetime (e.g., SARS, MERS, Ebola or Nipah)

## IV. MENTAL HEALTH AND PSYCHOLOGICAL SUPPORT (MHPSS)

The governments worldwide are taking preventing and containment measures for COVID-19, but also importance needs to be given on the mental and psychological wellbeing of citizens during such emergency crisis. The economic and social impacts of emergencies caused due to pandemic may be devastating and acute in the short term, which the countries eventually can cope up. On the other hand, an epidemic also threatens the long-term mental health and psychosocial wellbeing of the people, which often goes undermined and unnoticed. These impacts risk human rights, peace, and growth, as depicted in Table 2.

**Table 2: Mental health issues during emergency**

| Mental health issues | Social nature | Psychological nature |
|---|---|---|
| Pre Pandemic | • Extreme poverty;<br>• Belonging to a discriminated or marginalized group;<br>• Political oppression | • Mental disorder;<br>• Alcohol abuse |
| Pandemic induced | • Family separation;<br>• Disruption of social networks;<br>• Destruction of community structures, resources and trust;<br>• Increased gender-based violence | • Grief,<br>• Non-pathological distress;<br>• Depression and anxiety disorders,<br>• Post-traumatic stress disorder (PTSD)<br>• Suicidal tendencies |
| Humanitarian cause centred | • Undermining of community structures and traditional support mechanisms | • Anxiety due to a lack of basic necessity and food distribution information |

**Source:** WHO - IASC guidelines on mental health and psychosocial support in emergency settings

One of the priorities in emergencies is, therefore, to protect and improve people's mental health and psychosocial wellbeing. Emergency crisis creates a wide range of problems for the individual, family, community to society. The regular protective supports gets eroded, increasing the threat of diverse social problems which magnify already existing social injustice and inequality problems. Mental health problems compromises of both social and psychological issues. Inter-Agency Standing Committee (IASC) uses the term' *mental health and psychosocial support*' (MHPSS) in their Guidelines. They describe MHPSS in Emergency Settings as 'any type of local or outside support that aims to protect or promote psychosocial wellbeing an or/and prevent or treat mental health condition'. Also, the global humanitarian system uses the term MHPSS to unite a broad range of contributors responding to emergencies, including COVID-19 outbreak. These include people working in biological methods and sociocultural methods in social, education, health, and community settings. This underscores the need for diverse, complementary approaches in providing appropriate support. The impact of multifaceted humanitarian emergencies interventions on the mental health and psychosocial wellbeing of the population is multidimensional and endures long period after the crisis. While efforts to control and prevent the spread of the pandemic in the community are straight forward to follow, stereotyping and fear seem to have jeopardized efforts to respond (Ren et al. 2020). COVID-19 pandemic has already triggered hysteria; few examples are as demonstrated by a) bare toilet paper shelves in supermarkets, b) accusations against Asian(especially Chinese) communities (Malta et al.

2020), discriminating Asians and calling COVID-19 a Chinese virus, and e) impacted people's decisions to seek help when early symptoms arise (Ren et al. 2020).

**Fig 3. Pyramid of intervention for mental health and psychosocial support in emergencies.**

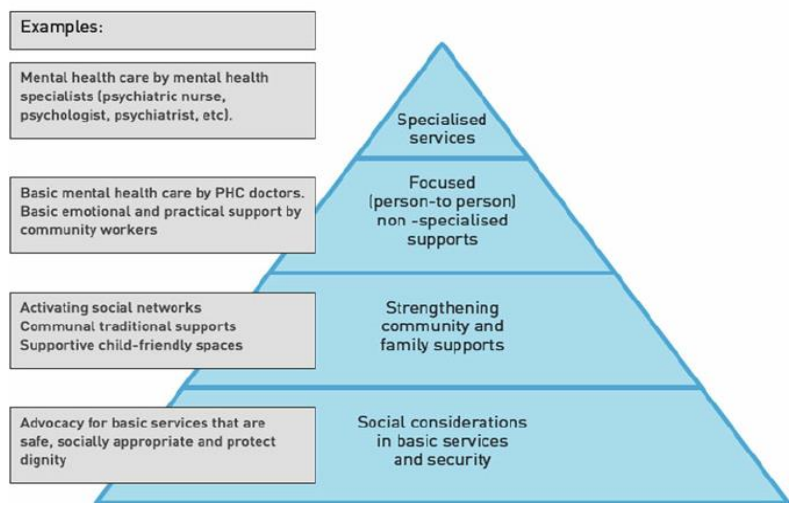

Although pandemic does not cause outright panic, it occurs as a result of large scale quarantine, social distancing and isolation in communities (Rubin & Wessely 2020). The current situation worldwide as a consequence of COVID-19 disease is unavoidable mass quarantines, which trigger mental and psychological illness. As people experience isolation, social distancing and mass quarantines for extended periods, this results in unmanageable extreme distress, anxiety and fear of losing control & getting trapped and surge in rumours getting spread (Rubin & Wessely 2020). The rumours feed irritability, uncertainty and are intertwined with social issues like compulsive hoarding and buying of good in an emergency, depression, fear, anxiety and psychological distress. Mental health professionals and psychologists speculate that COVID-19 pandemic can impact mental and psychological wellbeing of the population worldwide with the increase in the number of people affected of depression, self-harm, suicide, other than the symptoms stated globally due to coronavirus (Li et al.,2020; Moukaddam & Shah,2020; Yao et al.,2020).

## V. POSITIVE PSYCHOLOGY INTERVENTIONS

Martin Seligman in 1988 mainstreamed "positive psychology" by selecting it as the theme for his term as president of the American Psychological Association (Srinivasan, 2015; Tal., Ben-Shahar, 2007). The study of positive psychology arose to find reasons and answer for the mismatch between wealthy nations with growing economies, compared to their deteriorating life satisfaction, i.e. to the mental wellbeing of its citizens. With positive psychology came positive psychology interventions (PPI). PPIs are a combination of scientific strategies and tools which focus on increasing the wellbeing happiness, and positive emotions and cognitions among individuals (Keyes, 2002). PPIs are defined as psychological interventions that mainly focus on enhancing positive feelings, thoughts and behaviours, and comprises of two essential components; a) concentrate on happiness enhasement through positive thoughts and emotions; b) sustaining the outcomes of positive effects for long-term (Sin & Lyubomirsky, 2009). Park and Biswas-Dinner in 2013 proposed that for an intervention to be called PPI it must fulfil the following four requirements; a) Have a body of research supporting its reliability; b) Take into account one or more positive psychology constructs; c) Evidence-based & capable of being scientifically proven; and, d) benefit the practitioner lifelong.

PPIs are developed to promote positivity in individual's everyday life, and by doing so they support them to cope with the adverse events and negative moods they might experience (Seligman et al., 2006). PPIs have when applied on both clinically distressed and non-distressed population, have shown consistently positive results (Bolier et al., 2013; Stone and Parks, 2018). Individuals who have clarity of goals and expectations in life are more likely to experience happiness and contentment (Steger, Kashdan, & Oishi, 2008; Steger, Oishi, & Kashdan, 2009). Research suggests that practising of PPIs can serve as defensive mechanisms from developing mental disorders (Layous et al., 2014; Lyubomirsky & Layous, 2013). The amalgamation of enhanced wellbeing and several strategies employed though PPIs could allow for the reduction of mental and psychological risk factors for psychiatric disorders as rumination and loneliness; furthermore, PPIs help offset other possible environmental triggers through skills they develop (Layous et al., 2014). Thus, PPIs can be used in this COVID-19 pandemic to safeguard the mental and psychological wellbeing of general populations. PPIs are categories into seven types, (Park and Schueller, 2014), as discussed in table 2.

**Table 2: Positive Psychology Interventions types**

| Sl. No | Type of PPI | Core Focus | Outcomes |
|---|---|---|---|
| 1 | *Savouring* | Focus on a specific experience and aim to enhance their effects for maximizing happiness (Peterson, 2006). | Treating depression and mood disorders, produce happiness and self-satisfaction (Bryant, 2003). |
| 2 | Gratitude | Evoke strong feelings of positivity in the person who gives it and the person who receives it (Schueller & Parks, 2013). They are classified as: a) *Self-reflective practices* (e.g., writing a gratitude journal as too for self-expression; b) *Interactive methods:* The active expression of our gratitude to others (e.g., saying 'thank you,' giving small tokens of appreciation) | Increase happiness and satisfaction (Wood, Froh, & Geraghty, 2010). Makes one feel more positive and motivated from the inside (Emmons & McCullough, 2003; Seligman, Steen, Park, & Peterson, 2005). |
| 3 | Kindness | Unpretentious acts like buying someone a small token of love, offering for a noble cause, contributing something, or helping a stranger in need. (e.g., prosocial spending) | Promotes happiness-reinforcement and positivity (Howell and Iyer, 2012). |
| 4 | *Empathy* | Strengthen positive emotions in interpersonal relationships to develop healthy social bonds. (Diner & Seligman 2002).(e.g., self-love meditation, mindfulness practices (Fredrickson et al., 2008). | Happiness and inner peace. (Diener and Seligman, 2002). Building relations through broadened perception, effective communication, and bridge the gap between oneself and others. |
| 5 | Optimism | Create positive outcomes by setting realistic expectations. (e.g., a) Imagine Yourself test b) Life Summary technique) | Sense of positiveness about self and their life. (King, 2001). Gaining insight into where one is going wrong in life and what can be done to pursue the desired ideal life. |
| 6 | Strength | Focus on internal capacities and values (Parks and Biswas-Diener, 2013). (e.g., awareness and acknowledgement of power within) | Reduce depression and increases self-contentment (Seligman et al., 2005) |
| 7 | Meaning | Helps in understanding in life what is meaningful to us and why, and what can further be done to achieve the things that matter in life. (e.g., realizing meaning in our everyday activities, forming realistic goals and engaging effective means to attain them, or just reflecting on our emotions & thoughts (Grant, 2008). | Treat stress disorders, especially PTSD (Folkman and Moskowitz, 2000). |

**Source:** Author.

## VI. PPI'S AND MENTAL WELLBEING DURING COVID-19

PPIs have evidence-based literature of consistent positive outcomes for individuals mental and psychological wellbeing as discussed in table 2. PPIs enhance the psychological and mental wellbeing of individuals by reducing their stress, anxiety, fear, panic and symptoms of depression. These following PPIs can be self-learnt easily and practised on a day to day basis to tackle the mental anguish due to uncertainties of pandemics.

**Best Possible Self (BPS) :** A well-researched future-oriented intervention in positive psychology is BPS. First introduced by King (2001), the initial study demonstrated BPS had a significant effect on dispositional optimism and satisfaction with life and (King, 2001). This positive manipulation technique requires individuals to imagine themselves in an idea future, where everything has turned in their best favour, and where they have achieved all the dreams of their life. Individuals are required to write about this optimal future. Several studies in the past have explored the positive outcomes of imagining and writing practice of BPS, and the results have consistently, demonstrated that BPS increases and retains individual's positive mood, higher positive affect, psychological wellbeing and physical health (King, 2001; Peters et al., 2010; Sheldon & Lyubomirsky, 2006; Austenfeld, Paolo, & Stanton, 2006; Harrist, Carlozzi, McGovern, & Harrist, 2007). Further, results also indicate that imagining a positive future can certainly increase the expectancies for a positive future ( Madelon et al., 2010).

**Yoga and Mindfulness :** Yoga is rooted and practised in India from 5000 years ago (De-Michelis, 2005). Currently, in the west sprituality like yoga is increasingly practised as a way of life to successfully cultivate the aspects of overall psychological and mental wellbeing, which comprises a prime interest in positive psychology (Nimmi et al, 2021; Singleton, 2010). The term wellbeing encompasses self-actualization, optimal functioning and flourishing, and it concerns with both our desirable existence and the end stage of our pursuit (Ivtzan et al, 2013; Wong, 2011). Studies have demonstrated that practising yoga increases attention and awareness among individuals and lead to a mindful meditative state (Germer et al., 2005; Murphy & Donovan, 1997; Walsh, 1999). Yoga also results in an to increase of empathy experienced by individuals (Walsh, 2001), higher levels of compassion, awareness, gratitude, and respect toward ideas, beliefs and relationships (Radford, 2000). In literature, yoga intervention is strongly associated with stress reduction. The psychological mechanisms underlying ways through which yoga mitigates stress are, a) surge in a positive attitude towards stress (Woodyard et al., 2011), b) self-awareness (Arora and Battacharjee, 2008), c) coping mechanisms (Kinser, 2013), d) appraisal of control (Bonura, 2008), e) calmness (Sherman et al., 2013), f) spirituality (Evans et al., 2011), g) compassion (Braun et al., 2012), and h) mindfulness (Evans, 2011). Many studies suggest that the prevailing link between yoga practice and stress reduction is mindfulness (Dunn, 2008). Mindfulness is the state of being alert to and conscious of what is happening in the present (Brown and Ryan, 2003). Yoga practice ability to combat stress by being mindful has been widely established (Chiesa and Serretti, 2009). Any kind of yoga practise create a relaxing feeling to the body and mind and are considered as a practical and achievable technique to enhance psychological and mental wellbeing.

**Forgiveness: Self and Interpersonal :** Forgiveness is a useful tool for regulating negativity (McCullough et al. 1998). When people forgive, there is both a decrease of negative (angry and depressive) and an upsurge of positive and benevolent emotions, thoughts and behaviours towards the offending individual (Wade et al. 2014). Holding on to grudges and complaints hinder peace and prosperity. A beautiful positive psychology intervention that individuals can follow as a daily practice and imbibe into their personality is the art of forgiving. Mauger et al. (1992) classified forgiveness into two types according to the object of forgiveness: one is forgiveness toward others (interpersonal forgiveness) and the other is forgiveness toward one's self (self-forgiveness).  a) Interpersonal forgiveness: This is a series of process changes in social motivation when individuals have been offended; this type of forgiveness occurs when individuals replace destructive response with constructive behavior to achieve reconciliation with others (McCullough et al. 2001).  b) Self-forgiveness: This occurs when individuals part with their self-discontentment and provide compassion, tolerance, and love to themselves (Macaskill 2012; Xiao et al. 2014). Across all ages, both interpersonal and self forgiveness are associated with significantly lower rates of depression (Barcaccia et al. 2018;). Moreover, forgiveness is positively associated with wellbeing (Datu 2014; Pareek et al. 2016).

**Positive Affirmations :** Positive self-affirmations are 'having empathy for oneself'; these can be simple positive statements which indivials can say out load to themselves daily as a practice (eg: I deserve to be happy, I will love myself more from today), and fall into two categories (Armitage & Rowe, 2011). The first is attribute self affirmations which emphasis a person's positive attributes and skills. The second is value self affirmations which emphasize a persons' morals. Both kinds of self-affirmations boost positive behaviors (Fielden et al., 2016; Mays & Zhao, 2016), increase feelings of self-worth (Flynn and Chow, 2017), hopefulness & self esteem (Ivanoff and Ullrich, 2020), psychological well being (Emanuel et al., 2018) and enhance other wellbeing interventions (Howel, 2017). Positive self affirmations are statements of compliments, like "*verbal sunshine*" which bringsnd an immediate sense of pride and pleasure in the practing individual. Self affirmations as a PPI redirect the cognizance to focus on the positive sides in self and push individuals to act positively.

**Gratitude :** Gratitude is defned as both a positive afect ensuing from the perception of receiving a benefit from another individual (McCullough et al. 2002) and a trait, which comprises the ability to appreciate simple things in life, sense of abundance and experience and express gratitude towards others (Wood et al. 2009). When practiced, gratitude interventions have a positive effect on physical health and health behaviours (Boggins et al., 2020), enhancing sustainable mental health (Bohlmeijer et al., 2020) and wellbeing (Lai and  O'Carroll, 2017). Gratitude interventions like gratitude journaling, self-gratitude exercises and gratitude meditation, can easily be practiced by individuals to leverage their mental and psychological wellbeing.

**Psychological Capital (PsyCap) :** Psychological Capital is an important personal trait contributing to individual productivity by psychologists (Gohel, 2012). PsyCap is a second order construct and comprises of four dimensions, namely, self-efficacy, hope, optimism, and resilience (Luthans et al., 2008). As a personal resource, PsyCap positively influences the psychological success (Paul & Devi, 2018), psychological wellbeing (Mensah and Amponsah, 2016) and mitigates stress, anxiety and burnout (Demir, 2018). PsyCap can be easily learnt and practiced by individuals to strengthen their H.E.R.O elements (Hope, Self efficacy, Resilience, Optimism).

## VII. CONCLUSION

This study emphasizes and demonstrates that positive psychology interventions can be effective in the enhancement of subjective and psychological wellbeing. The research discussed various PPI which individuals can practice by on their own during the lockdown period in pandemic to nuture their mental and psychological well being. These PPI may help to reduce depressive symptom, stress, anxiety and hopelessness resulting from isolation and discrimination during COVID-19 pandemic. Mental health awareness in developing countries and efforts to bring them to light is a need of the hour, as already crashing economies mixed with the coronavirus pandemic can lead to a secondary crisis of mental wellbeing. Rapidly evolving technologies have made it feasible for the majority of individuals to gain more information on mental health and exercising PPI. Individualas, government authorities and NGOs are encouraged to take steps in their capacity to create awareness of psositive psychology and its benifts to combat the upcoming mental crisis if left ignored.

## VIII. ACKNOWLEDGEMENTS

We acknowledge all the frontline warrior's passion and service to humanity. We wholeheartedly thank them.

## REFERENCES

[1] Barcaccia, B., Pallini, S., Baiocco, R., Salvati, M., Saliani, A. M., & Schneider, B. H. (2018). Forgiveness and friendship protect adolescent victims of bullying from emotional maladjustment. *Psicothema*, *30*(4), 427-433.
[2] Boggiss, A. L., Consedine, N. S., Brenton-Peters, J. M., Hofman, P. L., & Serlachius, A. S. (2020). A systematic review of gratitude interventions: Effects on physical health and health behaviors. *Journal of Psychosomatic Research*, 110165.
[3] Bohlmeijer, E. T., Kraiss, J. T., Watkins, P., & Schotanus-Dijkstra, M. (2020). Promoting Gratitude as a Resource for Sustainable Mental Health: Results of a 3-Armed Randomized Controlled Trial up to 6 Months Follow-up. *Journal of Happiness Studies*, 1-22.
[4] Bolier, L., Haverman, M., Westerhof, G. J., Riper, H., Smit, F., & Bohlmeijer, E. (2013). Positive psychology interventions: a meta-analysis of randomized controlled studies. *BMC public health*, *13*(1), 119.
[5] Bryant, F. B. (2003). Savoring Beliefs Inventory (SBI): A scale for measuring beliefs about savoring. *Journal of Mental Health*, 12, 175-196
[6] Brooks SK, Webster RK, Smith LE, Woodland L, Wessely S, Greenberg N, et al. The psychological impact of quarantine and how to reduce it: rapid review of the evidence. *Lancet*. 2020;395:912–20.
[7] Centers for Disease Control and Prevention. (2020). Coronavirus Disease 2019 (COVID-19): People who are at higher risk for severe illness. Available from: https://www.cdc.gov/coronavirus/2019-ncov/specific-groups/high risk-complications.html
[8] Demir, S. (2018). The Relationship between Psychological Capital and Stress, Anxiety, Burnout, Job Satisfaction, and Job Involvement. *Eurasian Journal of Educational Research*, *75*, 137-153.
[9] Dubey S, Biswas P, Ghosh R, et al. Psychosocial impact of COVID-19. *Diabetes Metab Syndr*. 2020;14(5):779-788. doi:10.1016/j.dsx.2020.05.035
[10] Emanuel, A. S., Howell, J. L., Taber, J. M., Ferrer, R. A., Klein, W. M., & Harris, P. R. (2018). Spontaneous self-affirmation is associated with psychological wellbeing: Evidence from a US national adult survey sample. *Journal of Health Psychology*, *23*(1), 95-102.
[11] Emmons, R. A., & McCullough, M. E. (2003). Counting blessings versus burdens: An experimental investigation of gratitude and subjective wellbeing in daily life. *Journal of Personality and Social Psychology*, 84, 377-389
[12] Fielden, A.L., Little, L., Sillence, E., & Harris, P. (2016). Online self-affirmation increase fruit and vegetable consumptions in groups at high risk of low intake. *Applied Psychology: Health and Well-Being*, 8(1) 3-18. doi: 10.111/aphw.12059
[13] Flynn, D. M., & Chow, P. (2017). Self-efficacy, self-worth and stress. *Education*, *138*(1), 83-88.
[14] Gander, F., Proyer, R. T., & Ruch, W. (2016). Positive psychology interventions addressing pleasure, engagement, meaning, positive relationships, and accomplishment increase wellbeing and ameliorate depressive symptoms: A randomized, placebo-controlled online study. *Frontiers in psychology*, *7*, 686.
[15] Hendriks, T. (2018) : Positive psychology interventions in a multi-ethnic and cross-cultural context

[16] Holmes, E. A., O'Connor, R. C., Perry, V. H., Tracey, I., Wessely, S., Arseneault, L., & Ford, T. (2020). Multidisciplinary research priorities for the COVID-19 pandemic: a call for action for mental health science. *The Lancet Psychiatry*.

[17] Howell, A. J. (2017). Self-affirmation theory and the science of wellbeing. *Journal of Happiness Studies*, *18*(1), 293-311.

[18] IASC guidelines on mental health and psychosocial support in emergency settings. Geneva: Inter-Agency Standing Committee; 2007. Available from: http://www.who.int/mental_health/emergencies/9781424334445/en/

[19] Ivanoff, Stephanie and Ullrich, Taylor, "Hopefulness: Explaining the Link Between Self-Affirmation and Self-Esteem" (2020). *Psychology*. 9.

[20] Ji, W., Wang, W., Zhao, X., Zai, J., & Li, X. (2020). Crossspecies transmission of the newly identified coronavirus 2019-nCoV. *Journal of Medical Virology*, 92(4), 433–440.

[21] Kar S.K., Yasir Arafat S.M., Kabir R., Sharma P., Saxena S.K. (2020) Coping with Mental Health Challenges During COVID-19. In: Saxena S. (eds) Coronavirus Disease 2019 (COVID-19). *Medical Virology: From Pathogenesis to Disease Control*. Springer, Singapore

[22] Lai, S. T., & O'Carroll, R. E. (2017). 'The Three Good Things'-the effects of gratitude practice on wellbeing: a randomised controlled trial. *Health Psychol Update*, *26*, 10-18.

[23] Li, W., Yang, Y., Liu, Z. H., Zhao, Y. J., Zhang, Q., Zhang, L., Cheung, T., & Xiang, Y. T. (2020). Progression of Mental Health Services during the COVID-19 Outbreak in China. *International journal of biological sciences*, 16(10), 1732–1738. https://doi.org/10.7150/ijbs.45120

[24] Lyubomirsky, S., Sheldon, K. M., & Schkade, D. (2005) : Pursuing happiness -The architecture of sustainable change. *Review of General Psychology*, 9, 111-131.

[25] Mays, D., & Zhao, X. (2016). The influence of framed messages and self-affirmation on indoor tanning behavioral intention in 18- to 30-year old women. *Health Psychology*, 35, 123-130. doi: 10.1037/hea0000253.

[26] McCullough, M. E., Rachal, K. C., Sandage, S. J., Worthington, E. L., Wade Brown, S., & Hight, T. L. (1998). Interpersonal forgiving in close relationships: II. Theoretical elaboration and measurement. *Journal of Personality and Social Psychology*, 75, 1586–1603.

[27] McCullough, M. E., Emmons, R. A., & Tsang, J. A. (2002). The grateful disposition: A conceptual and empirical topography. *Journal of Personality and Social Psychology*, 82(1), 112–127.

[28] McCullough, M.E., Root, L.M., & Cohen, A.D. (2006) : Writing about the benefits of an interpersonal transgression facilitates forgiveness. *Journal of Consulting and Clinical Psychology*, 74, 887-897.

[29] Mensah, J., & Amponsah-Tawiah, K. (2016). Mitigating occupational stress: The role of psychological capital. *Journal of Workplace Behavioral Health*, *31*(4), 189-203.

[30] Mindfulness Tools For Happiness And Well-Being by EAP – https://humanresources.vermont.gov/sites/humanresources/files/Winter_2018_Messenger_State.pdf

[31] Moukaddam, N., & Shah, A. (2020). Psychiatrists beware! The impact of COVID-19 and pandemics on mental health. *Psychiatric Times*, 37(3). https://www.psychiatrictimes.com/psychiatrists-bewareimpact-coronavirus-pandemics-mental-health

[32] Nimmi, P.M., Binoy, A.K., Joseph, G. and Suma, R. (2021), "Significance of developing spirituality among management students: discerning the impact on psychological resources and wellbeing", *Journal of Applied Research in Higher Education*, Vol. ahead-of-print No. ahead-of-print. https://doi.org/10.1108/JARHE-10-2020-0372

[33] Paul V, M.T., Aboobaker, N. and Devi, N, U. (2021), Family incivility, burnout and job satisfaction: examining the mediation effect, *Benchmarking: An International Journal*, Vol. ahead-of-print No. ahead-of-print. https://doi.org/10.1108/BIJ-10-2020-0534

[34] Paul, V. M. T., & Devi, N. U. (2018). Exploring the relationship between psychological capital and entrepreneurial success. *International Journal of Pure and Applied Mathematics*, 119(18), 2987-2999.

[35] Pawelski, J. O. (2020). The elements model: toward a new generation of positive psychology interventions. *The Journal of Positive Psychology*, *15*(5), 675-679.

[36] Peterson, C. (2009). Positive psychology. *Reclaiming children and youth*, *18*(2), 3.

[37] Qiu, J., Shen, B., Zhao, M., Wang, Z., Xie, B., & Xu, Y. (2020). A nationwide survey of psychological distress among Chinese people in the COVID-19 epidemic: implications and policy recommendations. *General psychiatry*, *33*(2).

[38] Ren, S.-Y., Gao, R.-D. & Chen, Y.-L. (2020). Fear can be more harmful than the severe acute respiratory syndrome coronavirus 2 in controlling the corona virus disease 2019 epidemic. *World Journal of Clinical Cases*, 8 (4), 652–657.

[39] Seligman, M. E. (2004). *Authentic happiness: Using the new positive psychology to realize your potential for lasting fulfillment*. Simon and Schuster.
[40] Seligman, M. E., & Csikszentmihalyi, M. (2014). Positive psychology: An introduction. In *Flow and the foundations of positive psychology* (pp. 279-298). Springer, Dordrecht.
[41] Shigemura J, Ursano RJ, Morganstein JC, Kurosawa M, Benedek DM. Public responses to the novel 2019 coronavirus (2019-nCoV) in Japan: mental health consequences and target populations. *Psychiatry Clin Neurosci*. 2020 [Epub ahead of print].
[42] Sin, N. L., & Lyubomirsky, S. (2009). Enhancing wellbeing and alleviating depressive symptoms with positive psychology interventions: A practice-friendly meta-analysis. *Journal of clinical psychology*, *65*(5), 467-487
[43] Slade, M. (2010). Mental illness and wellbeing: the central importance of positive psychology and recovery approaches. *BMC health services research*, *10*(1), 1-14.
[44] Stone, B. M., & Parks, A. C. (2018). Cultivating subjective wellbeing through positive psychological interventions. *Handbook of Well-being*. Srinivasan, T. S. (2015). The 5 Founding Fathers and A History of Positive Psychology.
[45] Tal., Ben-Shahar (2007). *Happier : learn the secrets to daily joy and lasting fulfillment*. New York: McGraw-Hill. ISBN 978-0071510967.
[46] Varshney M, Parel JT, Raizada N, Sarin SK (2020) Initial psychological impact of COVID-19 and its correlates in Indian Community: An online (FEEL-COVID) survey. *PLoS ONE* 15(5): e0233874. https://doi.org/10.1371/journal.pone.0233874
[47] Vázquez, C., Hervás, G., Rahona, J. J., & Gómez, D. (2009). Psychological wellbeing and health. Contributions of positive psychology. *Annuary of Clinical and Health Psychology*, *5*(2009), 15-27
[48] Wade, N. G., Hoyt, W. T., Kidwell, J. E., & Worthington, Jr., E. L. (2014). Efficacy of psychotherapeutic interventions to promote forgiveness: a meta-analysis. *Journal of Consulting and Clinical Psychology*, 82(1), 154–170.
[49] Wang, W., Tang, T. & Wei, F. (2020). Updated understanding of the outbreak of 2019 novel coronavirus (2019-nCoV) in Wuhan, China. *Journal of Medical Virology*, 92 (4), 441–447.
[50] Wood, A. M., Joseph, S., & Maltby, J. (2009). Gratitude predicts psychological wellbeing above the Big Five facets. *Personality and Individual Diferences*, 46(4), 443–447.
[51] World Health Organization. (2018). Mental Health ATLAS 2017. https://www.who.int/mental_health/evidence/atlas/profiles-2017/ IND.pdf?ua=1
[52] World Health Organization. (2020a). COVID-19 and violence against women: What the health sector/system can do. WHO, 26 March 2020. https://www.who.int/reproductivehealth/publications/emergencies/COVID-19-VAW-full-text.pdf
[53] World Health Organization. (2020b). Helping children cope with stress during the 2019-nCoV outbreak. https://www.who.int/docs/defaultsource/coronaviruse/helping-children-cope-with-stress-print.pdf?sfvrsn=f3a063ff_2&ua=1
[54] World Health Organization. (2020c). Mental health and COVID-19. http://www.euro.who.int/en/health-topics/health-emergencies/coronavirus-covid-19/novel-coronavirus-2019-ncov-technical-guidance/ coronavirus-disease-covid-19-outbreak-technical-guidance-europe/ mental-health-and-covid-19
[55] World Health Organization. (2020d). Mental health and psychosocial considerations during the COVID-19 outbreak. WHO reference number: WHO/2019-nCoV/MentalHealth/2020.1. https://www.who.int/docs/default-source/coronaviruse/mental-health-considerations.pdf
[56] Yao. H., Chen, J., & Xu, Y. (2020). Patients with mental health disorders in the COVID-19 epidemic. *The Lancet*, 7(4), e21. https://doi. org/10.1016/S2215-0366(20)30090-0
[57] Yu-Tao X, Yang Y, Li W, Zhang L, Zhang Q, Cheung T,et al. Timely mental health care for the 2019 novel coronavirus outbreak is urgently needed. *Lancet Psychiatry*. 2020;7:228–9.